\documentclass[aps,prb,preprint,amsfonts,showpacs,letterpaper]{revtex4-1}

\usepackage{CJK} 
\usepackage{graphicx, multirow,bm}
\usepackage{hyperref}

\begin{document}
\begin{CJK*}{UTF8}{bsmi} 
	\title{Self-consistent density functional calculations of the crystal field levels in lanthanide and actinide dioxides}
	\author{Fei Zhou(周非)}
	\author{Vidvuds Ozoli\c{n}\v{s}} 
	\affiliation{Department of Materials Science and Engineering, University of California, Los Angeles, CA 90095, USA} 
	\date{\today} 
	\begin{abstract}
Using a recently developed method combining a nonspherical self-interaction corrected LDA+$U$ scheme and an on-site multi-body Hamiltonian [Phys.\ Rev.\ B 83, 085106 (2011)], we calculate the crystal field parameters and crystal field (CF) excitation levels of $f$-element dioxides in the fluorite structure with $f^{n}$ electronic configurations, including $n=1$ (PaO$_{2}$, PrO$_{2}$), $n=2$ (UO$_{2}$), $n=3$ (NpO$_{2}$), and $n=4$ (PuO$_{2}$). It is shown that good agreement with experimental data (within approximately 10 to 20 meV) can be obtained in all cases. The properties of the multi-electron CF ground states are analyzed.
\end{abstract}
\maketitle
\end{CJK*}

\section{Introduction}


The electronic structure of lanthanide and actinide compounds has a number of distinctive features that are manifestations of atomic $f$-electron physics in bulk solids, including strong on-site correlations and relativistic spin-orbit effects. The effects of chemical environment on the ground states and excitation spectra of $f$-electrons are particularly interesting, since they are responsible for splitting the otherwise $2J+1$-fold degenerate free-ion ground state $^{2S+1}L_{J}$, giving rise to rich physics and applications. 
Recently, actinide dioxides in the cubic fluorite structure have attracted renewed theoretical interest in the context of their use as nuclear fuels \cite{Petit2010PRB45108, Nakamura2010PRB155131,Zhang2010PRB144110, Meredig2010PRB195128, Alexandrov2010PRB174115, Suzuki2010PRB241103, Andersson2009PRB60101, Dorado2009PRB235125,Tiwary2009PRB174302, Geng2008PRB180101}.


The crystal field (CF) method is a well established tool for describing the ligand environment of localized electrons. In its conventional form, which requires spectroscopic information for fitting the CF parameters, the CF method has been applied to $f$-electron compounds with considerable success \cite{Wybourne1965Spectroscopic,  Newman2000, Liu2005Spectroscopic}, including numerous characterizations of actinide oxides.\cite{Rahman1966JPCS1833, Amoretti1989PRB1856, Amoretti1992JPCM3459, Kern1999PRB104,Gajek2004JMMM415,Magnani2005PRB54405, Magnani2007JPCS2020, Santini2000PRL2188, Kern1984SSC295, Boothroyd2001PRL2082, Magnani2008PRB104425}  Using first-principles density functional theory (DFT) approaches, Divis and co-workers calculated the crystal field in praseodymium oxides \cite{ Divis2005JMMM1015,Novak2007PSSB3168}, and Colarieti-Tosti and co-workers studied PuO$_{2}$ \cite{Colarieti-Tosti2002PRB195102}.   
However, since the CF splitting is much weaker that the Coulomb repulsion and spin-orbit coupling (SOC), DFT-based calculations are often plagued by various technical issues, such as lack of a fully self-consistent treatment of the $f$-charge density or explicit consideration of electronic correlation. Recently Gaigalas and co-workers \cite{Gaigalas2009LJP403} calculated CF levels of actinide dioxides with relativistic quantum chemical methods.

We have recently developed a fully self-consistent method of calculating the CF parameters, which combines an improved nonspherical self-interaction free LDA+$U$ scheme \cite{Zhou2009PRB125127} with a model on-site Hamiltonian including Coulomb, spin-orbit, and CF terms.\cite{Zhou2011PRB85106} Our approach utilizes the existence of multiple local minima in the LDA+$U$ total energy functional and uses the corresponding $f$-electron wavefunctions and total energies to extract CF parameters. Good agreement with experiment was obtained in terms of the predicted UO$_{2}$ CF excitation spectrum (within about $10$ to $20$ meV) and magnetic properties of UO$_{2}$.\cite{Zhou2011PRB85106} In this paper, we extend this method to calculate the CF parameters of other $f$-element dioxides MO$_{2}$ in the fluorite structure with the $f^{n}$ configuration, including $n=1$ (PaO$_{2}$, PrO$_{2}$), $n=3$ (NpO$_{2}$), and $n=4$ (PuO$_{2}$). Some results for UO$_{2}$ ($n=2$) are included for completeness. Other $f$-elements are not considered either because they have no valence $f$ electrons (CeO$_{2}$, ThO$_{2}$), no stable dioxides (heavier lanthanides), or no suitable pseudopotential presently available to us (AmO$_{2}$, CmO$_{2}$).

\section{Method}
The CF of MO$_{2}$ in the fluorite structure is given by:
\begin{eqnarray}
H_{\mathrm{CF}}&=& \frac{16\sqrt{\pi}}{3}  V_{4} (Y_{4}^{0} + \sqrt{\frac{10}{7}} \mathfrak{Re} Y_{4}^{4}) \nonumber \\
 &+&   32 \sqrt{\frac{\pi}{13}} V_{6} (Y_{6}^{0} - \sqrt{14} \mathfrak{Re} Y_{6}^{4}), \label{eq:CF} 
\end{eqnarray}
where $V_{4}$ and $V_{6}$ are CF parameters of the cubically coordinated metal ion, related to other common CF notations \cite{Newman2000} by
\begin{eqnarray*}
V_{4}&=& B_{4}/8, \\ 
V_{4}&=& B_{6}/16. 
\end{eqnarray*}
In addition, free-ion parameters $F^{k}$ ($k=2,4,6$) and $\zeta$ describe the Coulomb and SOC terms, respectively, in the total Hamiltonian:
\begin{eqnarray}
	H &=& H_{\mathrm{CF}} + \hat{V}_{\mathrm{ee}} +  \zeta \hat{\boldsymbol l} \cdot \hat{\boldsymbol  s}.
 \label{eq:CI-Hamiltonian} 
\end{eqnarray}
where $\hat{V}_{\mathrm{ee}}$ designates the Coulomb repulsion between $f$-electrons. Since the Slater integrals $F^{k}$ ($k=2,4,6$) in $\hat{V}_{\mathrm{ee}}$ are heavily correlated \cite{Carnall1992JCP8713}, the following approximation has been adopted:\cite{Berry1988CP105} 
\begin{eqnarray}
F^2&=&F^4/0.668=F^6/0.494,  
\label{eq:Fk-to-J}
\end{eqnarray}
eliminating free parameters  $F^{4}$ and $F^{6}$. The Hamiltonian of Eq.~(\ref{eq:CI-Hamiltonian}) is diagonalized with $f^{n}$ basis wavefunctions, which are chosen in this work as $n$-body Slater determinants constructed from 14 $f^{1}$ spin-orbitals $\{   Y_{3}^{m} \sigma \}$ ($m=-3 ,\cdots ,3$, $\sigma=\uparrow, \downarrow$). Therefore there are $C^{n}_{14}$ basis wavefunctions to expand an $f^{n}$ state.

\begin{table}[htbp] 
	\begin{ruledtabular}
		\begin{tabular}{|c|ccccc|} 
		&PrO$_{2}$	&PaO$_{2}$	&UO$_{2}$	&NpO$_{2}$	&PuO$_{2}$\\ 
		\hline
$a$ (\AA)\cite{Villars1991PearsonHandbook}	&5.386 		&5.505 	&5.470		&5.433 	&5.396 \\
$f$-conf.		& $f^{1}$			&$f^{1}$			&$f^{2}$			&$f^{3}$			&$f^{4}$\\
Ion GS. 	&$^{2}F_{5/2}$	&$^{2}F_{5/2}$	&$^{3}H_{4}$	&$^{4}I_{9/2}$	&$^{5}I_{4}$ \\
CF GS		&$\Gamma_{8}$(4)	&$\Gamma_{8}$(4)	&$\Gamma_{5}$(3)	&$\Gamma_{8}^{(1)}$(4)	&$\Gamma_{1}$(1)	\\
$J$ (eV)		&0.78	&0.58	& 0.60	& 0.58	& 0.55	\\
		\end{tabular}
	\end{ruledtabular}
	\caption{List of studied $f$-element oxides MO$_{2}$ (M=Pr, Pa, U, Np, Pu), including the lattice constants $a$, number of localized $f$-electrons $n$, free-ion and crystal field ground states (GS) and their multiplicities (in parentheses), and the  $J$ parameter used in LDA+$U$ calculations.\cite{Zhou2009PRB125127}
	\label{tab:input}} 
\end{table}

All DFT calculations were carried out using the same computational settings as in our previous work \cite{Zhou2011PRB85106}. Input parameters for the LDA+$U$ \cite{Anisimov1991PRB943, Zhou2009PRB125127} corrections are chosen as $U=6$ eV, $c=0.5$ and the $J$ parameter for exchange interactions are determined by the requirement of numerical degeneracy of degenerate ionic states \cite{Zhou2009PRB125127}. For each compound, 50 calculations with randomly initialized $f^{n}$ wavefunctions were carried out at the experimental lattice parameters (Table~\ref{tab:input}). The magnetization axis for analyzing the energy eigenstates is chosen along $z$. More details of our technical approach can be found in Ref.~\onlinecite{Zhou2011PRB85106}.

\section{Results and discussions}
The calculated model parameters are summarized in Table~\ref{tab:summary}. For comparison, we also give the free-ion parameters $F^{k}$ and $\zeta$ of tetravalent actinides in the corresponding fluorides.\cite{Carnall1989JLM201}
The Coulomb interactions $F^{k}$ do not enter the Hamiltonian of the $f^{1}$ compounds PrO$_{2}$ and PaO$_{2}$. $F^{k}$ is found to be slightly smaller in NpO$_{2}$ than UO$_{2}$ and PuO$_{2}$, in agreement with the trend observed in MF$_{4}$. The SOC parameters $\zeta_{5f} \approx 0.2$ to $0.3$ eV of the heavier actinides are found substantially larger than  lanthanide, since relativistic effects are more pronounced in heavier elements. The calculated $\zeta_{5f}$ of PaO$_{2}$ is almost twice as large as the corresponding $\zeta_{4f}=0.115$ eV in the rare earth compound PrO$_{2}$.  $\zeta$ is predicted to increase over the actinide series, in agreement with experiment. However, our calculated $\zeta_{5f}$ values are overestimated by $5$ to $15\%$. 


Higher localization of the $4f$ states explains the smaller CF parameters $V_{4}$ and $V_{6}$ in PrO$_{2}$ compared to $5f$ actinides. The 4th-order CF parameter $V_{4}$ is significantly larger than $V_{6}$ for all the dioxides, in agreement with results obtained from fitting experimental spectra \cite{Amoretti1989PRB1856,Magnani2005PRB54405}. 

\begin{table}[htbp] 
	\begin{ruledtabular}
		\begin{tabular}{|c|ccccc|} 
		&PrO$_{2}$	&PaO$_{2}$	&UO$_{2}$	&NpO$_{2}$	&PuO$_{2}$\\ 
		\hline
$F^{2}$		&	&	&5.649	& 5.004	&6.147	\\
$\zeta$		&0.115	& 0.210	&0.230	&0.293	&0.304	\\
$V_{4}$		&-0.067	& -0.113	&-0.093	&-0.082	&-0.099	\\
$V_{6}$		&0.005	& 0.015	&0.016	&0.014	&0.017	\\ 
& \multicolumn{5}{c|}{Free-ion parameters in MF$_{4}$ from Ref.~\onlinecite{Carnall1989JLM201}} \\
$F^{2}$		&	&	&5.86	&5.55	&5.88	\\
$\zeta$		& 	&  	&0.22	&0.25	&0.28	\\
\hline
& \multicolumn{5}{c|}{Excited CF levels} \\
State		& $\Gamma_{7}(2)$		& $\Gamma_{7}(2)$		& $\Gamma_{3}(2)$		&$\Gamma_{8}^{(2)}(4)$		&$\Gamma_{4}(3)$ \\
		&		&		&$\Gamma_{4}(3)$, $\Gamma_{1}(1)$			& $\Gamma_{6}(2)$		& $\Gamma_{3}(2)$, $\Gamma_{5}(3)$ \\
Pred.	& 0.129	&0.186	&0.126, 	&0.034	&0.097	\\
		&		&		&0.158, 0.176 			& 0.125		& 0.195, 0.204 \\
		\cline{2-6}
Expt.		&0.131	& n/a 	&0.150,	& 0.055	& 0.123	\\
		&		&		& 0.158, 0.170			& n/a		&  n/a  \\
Ref.		&\onlinecite{Kern1984SSC295}, \onlinecite{Boothroyd2001PRL2082}		&		& \onlinecite{Nakotte2010JPCS12002}			& \onlinecite{Amoretti1992JPCM3459}		& \onlinecite{ Kern1999PRB104} \\ \cline{2-6} 
Calc.		&  	&   	&  	&   	& 0.099	\\
(Ref.~\onlinecite{Colarieti-Tosti2002PRB195102})		&		&		&  			&  	& 0.162, 0.208   \\  \cline{2-6} 
Calc.		& 0.082	&  	& 0.167	& 0.056	& 0.112	\\
(Ref.~\onlinecite{Magnani2005PRB54405})		&	 	&		& 0.187, n/a			& n/a		&  n/a  \\ \cline{2-6}
Calc.		& 	&   	&0.155	& 0.034	& 0.064	\\
(Ref.~\onlinecite{Gaigalas2009LJP403})		&		&		& 0.161, 0.189			& 0.099	&  0.103, 0.127  \\ \cline{2-6} 
	\end{tabular}
	\end{ruledtabular}
	\caption{Calculated model parameters and energy eigenvalues in eV. $F^{4,6}$ may be derived from eq.~\ref{eq:Fk-to-J}. Excited CF levels are labeled with the corresponding degeneracy in parentheses and compared with reported measured or calculated values.
	\label{tab:summary}} 
\end{table}

\begin{figure}[htbp] 
	\includegraphics[width=0.99 \linewidth]{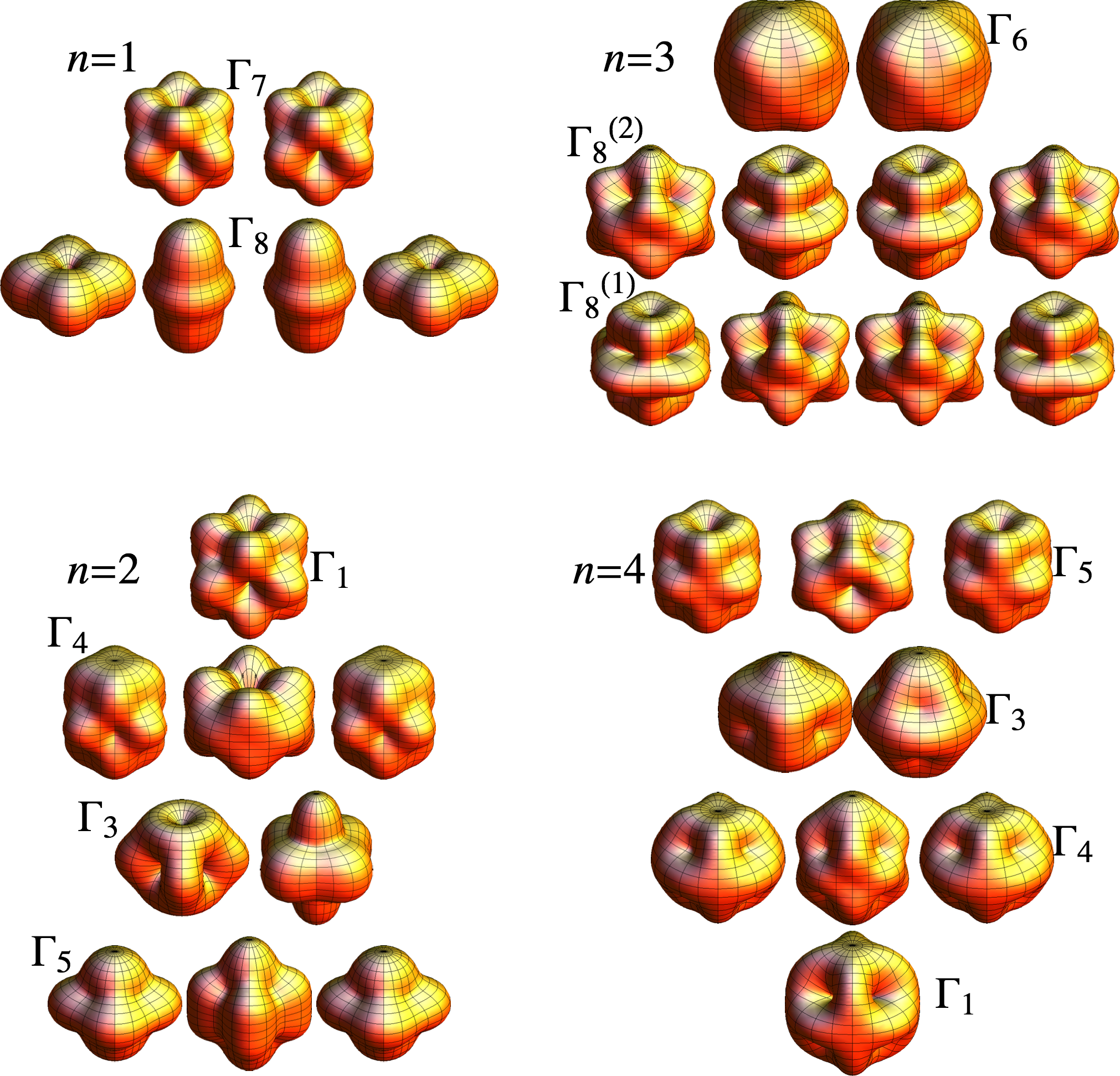} 
	\caption{The $f^{n}$ ($n=1,\cdots, 4$) eigenstates under the fluorite cubic crystal field. Only the $2J+1$ lowest states are shown.} \label{fig:eigenstates} 
\end{figure}

Using the parameters given in Table~\ref{tab:summary}, the  crystal field eigenstates are obtained by diagonalizing the effective Hamiltonian in Eq.~\ref{eq:CI-Hamiltonian}; the resulting wave functions are visualized in Fig.~\ref{fig:eigenstates}. Following the procedure of Refs.~\onlinecite{Sievers1982ZPB289,*Walter1986ZPB299}, the radius $R(\Omega=\theta,\phi)$ of the spherical plots of the charge distribution is 
$$
R(\Omega)= (\rho(\Omega)- \bar\rho )^{1/3},
$$
where $\rho(\Omega)$ is the spherical part of the charge distribution centered at the metal ion, and $\bar\rho$ is an appropriate amount of monopole subtracted from $\rho(\Omega)$ to emphasize its asymmetric character.

The predicted low-energy CF excitation levels are shown in Table~\ref{tab:summary} alongside available experimental data. Qualitatively, the correct ground states and ordering of the excited states are obtained in all cases. Quantitatively, good agreement with the measured spectrum has been obtained, with the errors in the excitation energies being within 10 to 20 meV. Reasonable agreement has also been found with previous theoretical calculations reported in Refs.~\onlinecite{Colarieti-Tosti2002PRB195102, Magnani2005PRB54405, Gaigalas2009LJP403}. Next, we discuss each compound in detail (except UO$_{2}$).

\subsection{PrO$_{2}$ and PaO$_{2}$}
The $f^{1}$ compounds PrO$_{2}$ and PaO$_{2}$ differ from $n>1$ cases in that the multiple local-minima issues that plague calculations for multi-$f$ electron systems are less severe. Out of the 50 random wave function initializations, approximately 5\% with highly unstable starting states failed to converge within 100 electronic steps and were discarded. The rest exhibited a relatively small energy spread and were all within $0.2$ to $0.3$ eV from the CF ground state, compared to the spread of about 2 eV observed in UO$_{2}$.\cite{Zhou2011PRB85106} This shows that the many-body interaction is a main reason for the existence of many local-minimum solutions, and without this obstacle the $f^{1}$ calculations can find the $j=5/2$ Russell-Sanders ground state, even though they may fail in locating the CF ground state. The original LDA+$U$ \cite{Anisimov1991PRB943} scheme was tested for PrO$_{2}$ and PaO$_{2}$ and found to increases significantly the energy spread of the local-minimum solutions due to orbital-dependent self-interaction errors \cite{Zhou2009PRB125127}. 

The predicted $\Gamma_{8} \rightarrow \Gamma_{7}$ excitation energy for PrO$_{2}$ is 129 meV, in excellent agreement with the measured value of 131 meV \cite{Boothroyd2001PRL2082} and more accurate than our previous rough estimation of $73$ to $142$ meV in Ref.~\onlinecite{Zhou2009PRB125127}, showing that our method based on Eq.~(\ref{eq:CI-Hamiltonian}) leads to significant error cancellation in the calculated CF energies. Predictions for the higher CF levels of $J=7/2$ are 0.376 ($\Gamma'_{6}$), 0.433 ($\Gamma'_{8}$) and 0.622 eV ($\Gamma'_{7}$), respectively, compared with observed values of 0.320, 0.390 and 0.580 eV from Ref.~\onlinecite{Boothroyd2001PRL2082}. The relative splitting within the $J=7/2$ manifold agrees very well with experiment, showing the validity of our predicted CF parameters, while the center of these levels are 11\% too high, due to the over-estimated spin-orbit coupling (our $\zeta=0.115$ eV compared to 0.1 eV of Ref.~\onlinecite{Boothroyd2001PRL2082}). Experiments on PaO$_{2}$ are relatively scarce. The only available number of 140 meV for the $\Gamma_{8} \rightarrow \Gamma_{7}$ transition cited in Ref.~\onlinecite{Konings2004JCT121} is based on private communications, which we inquired about but could not confirm. Our prediction of 186 meV for $\Gamma_{8} \rightarrow \Gamma_{7}$ in PaO$_{2}$ is substantially larger than the corresponding value for PrO$_{2}$, in agreement with the trends in CF parameters in Table~\ref{tab:summary}.

\begin{table}[tbp]
\newlength{\graphheight} 
\graphheight 0.6cm
	\begin{ruledtabular}
		\begin{tabular}{|l|cc|cc|}
\multicolumn{1}{|c|}{State} & \multicolumn{2}{c|}{PrO$_{2}$}  & \multicolumn{2}{c|}{PaO$_{2}$}  \\
 &$\mu_{S}$ & $\mu$  &$\mu_{S}$ & $\mu$   \\ \hline
 (1) $\Gamma_{8}$ \includegraphics[height=\graphheight]{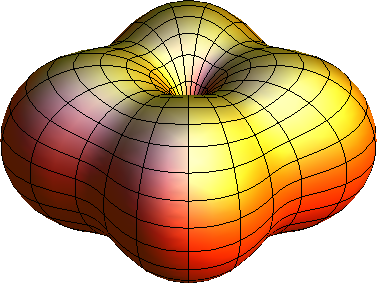}& -0.64& 1.49 & -0.57 & 1.54  \\
 (2) $\Gamma_{8}$ \includegraphics[height=\graphheight]{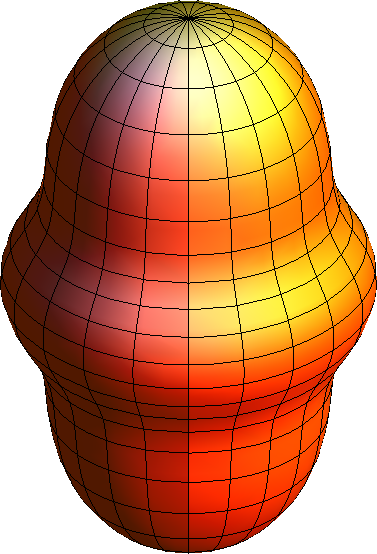}& -0.03 & 0.47  & -0.10 & 0.45 \\
 (3) $\Gamma_{8}$ \includegraphics[height=\graphheight]{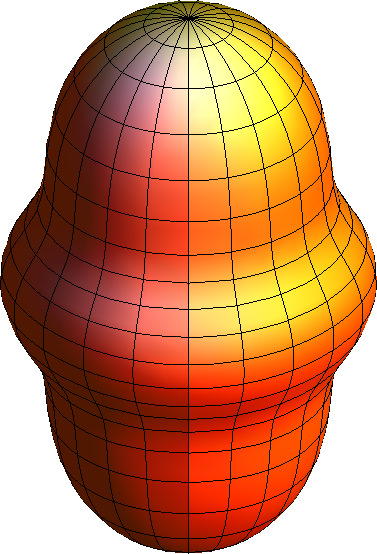}& 0.03 & -0.47   & 0.10 & -0.45\\ 
 (4) $\Gamma_{8}$ \includegraphics[height=\graphheight]{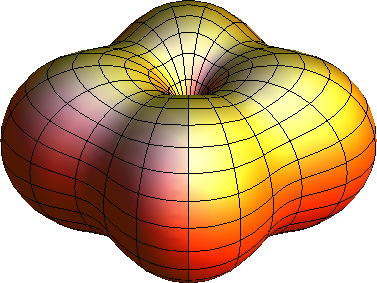}& 0.64 & -1.49  & 0.57 & -1.54  \\
 (5) $\Gamma_{7}$ \includegraphics[height=\graphheight]{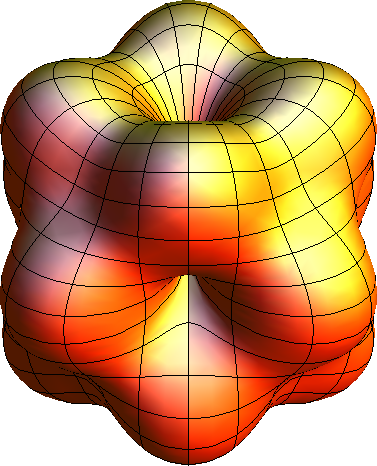}& 0.06 & 0.73   & 0.10 & 0.70\\ 
 (6) $\Gamma_{7}$ \includegraphics[height=\graphheight]{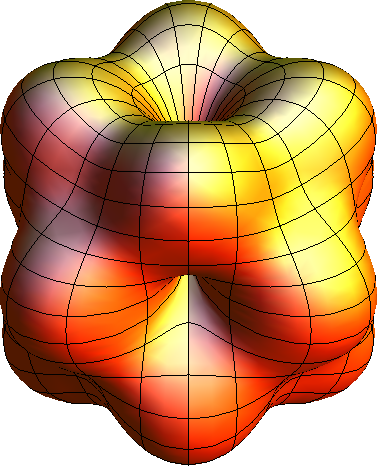}& -0.06 & -0.73  & -0.10 & -0.70  \\
		\end{tabular}
	\end{ruledtabular}
	\caption{The $\Gamma_{8}$ ground state quartet and $\Gamma_{7}$ excited doublet of PrO$_{2}$ and PaO$_{2}$ ($f^{1}$) in the fluorite structure, and the corresponding spin and total magnetic moment in $\mu_{B}$.}
	\label{tab:f1} 
\end{table}

\subsection{UO$_{2}$}
Recent measurement by \textcite{Nakotte2010JPCS12002} of the crystal field levels in UO$_{2}$ provides updated information than \textcite{Amoretti1989PRB1856}: the excitation peak at 180 meV is spurious. 

\subsection{NpO$_{2}$} 
NpO$_{2}$ has the $5f^{3}$ configuration. The excitation energy between the CF ground state $\Gamma_{8}^{(2)}$ and the first excited $\Gamma_{8}^{(1)}$ state has been measured to be 55 meV \cite{Amoretti1992JPCM3459}. Our prediction of 34 meV is a reasonable under-estimation. Note that a recent quantum chemical calculation \cite{Gaigalas2009LJP403} for NpO$_{2}$ predicted the same value as ours. Two estimated values for the second excited $\Gamma_{6}$ energy level (145 meV and 274 meV) are given in Ref.~\onlinecite{Amoretti1992JPCM3459},  and only the first value scales over the actinide dioxides series \cite{Magnani2005PRB54405}. We predict an excitation energy of 125 meV for $\Gamma_{8}^{(2)} \rightarrow \Gamma_{6}$, which agrees with the latter assessment.

\begin{table}[tbp]
\graphheight 0.6cm
	\begin{ruledtabular}
		\begin{tabular}{|c|ccc|}
State &$\mu_{S}$ & $\mu$  & Proj.   \\ \hline
 $a$ \includegraphics[height=\graphheight]{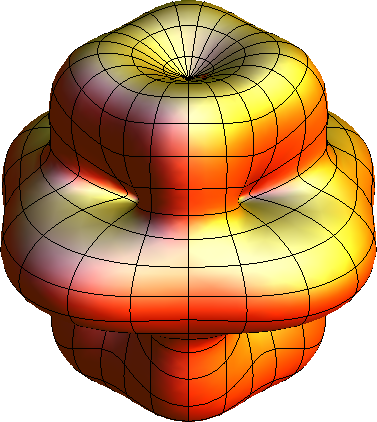}& -1.00& 1.72 & $0.68(1,2,3)+0.57(1,5,6)   $\\
 $b$ \includegraphics[height=\graphheight]{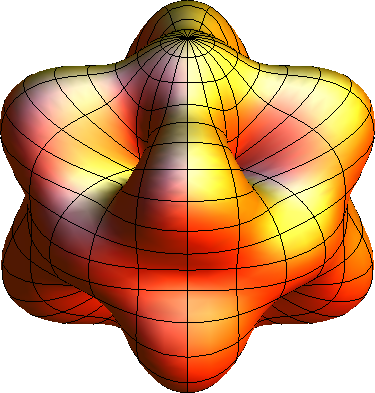}& -0.37 & 0.57  & $0.68(1,2,4)+0.57(2,5,6)  $\\
 $c$ \includegraphics[height=\graphheight]{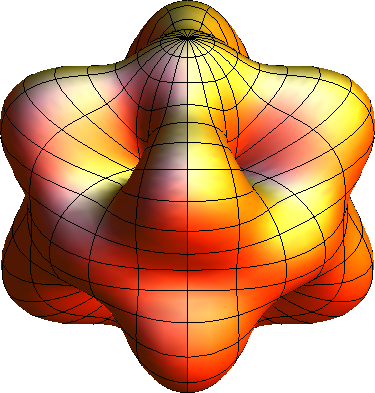}& 0.37 & -0.57   & $0.68(1,3,4)+0.57(3,5,6) $\\ 
 $d$ \includegraphics[height=\graphheight]{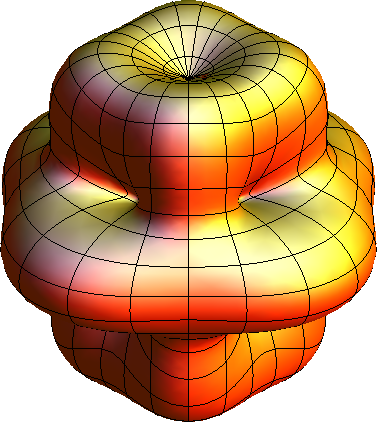}& 1.00 & -1.72  &  $0.68(2,3,4)+0.57(4,5,6) $\\
		\end{tabular}
	\end{ruledtabular}
	\caption{The $\Gamma_{8}^{(2)}$ ground state quartet of NpO$_{2}$ ($5f^{3}$), along with their magnetic moment and projection onto Slater determinants (only a few leading terms shown) composed of $f^{1}$ eigenstates $\Gamma_{8}$ and $\Gamma_{7}$. One-electron orbitals (1) through (6) are defined in Table~\ref{tab:f1}.}
	\label{tab:f3} 
\end{table}

Since the $f^{1}$ configuration in the fluorite structure is split by the crystal field into the $\Gamma^{8}$ quartet and $\Gamma_{7}$ doublet (see Table~\ref{tab:f1}), the multi-electron configurations $f^{n}$ of UO$_{2}$, NpO$_{2}$ and PuO$_{2}$ are sometimes interpreted within a picture where the added electrons gradually fill the CF levels, in analogy to the well-known scenario of $d$-electrons filling the $t_{2g}$ and $e_{g}$ CF levels in transition metal compounds. As we have shown previously \cite{Zhou2011PRB85106}, this picture fortuitously holds for the $\Gamma_{5}$ ground state of UO$_{2}$. However, transition metal CF splittings are usually several eV, while for $f$ the CF splittings (on the order of 0.1 eV) is much smaller than the effective Coulomb interactions ($\sim$ eV). Hence, the $f^{n}$ eigenstates are in general multi-configurational. According to Table \ref{tab:f3}, the $\Gamma_{8}^{(2)}$ ground states of NpO$_{2}$ are composed of multiple determinants, including ones with substantial projections onto not only the $\Gamma_{8}$ CF ground states, but also the $\Gamma_{7}$ excited states of $f^{1}$. In other words, the $f^{3}$ ground state $\Gamma_{8}^{(2)}$ occupies both $\Gamma_{8}$ and $\Gamma_{7}$ orbitals in order to lower its electrostatic energy at the expense of a slightly increased CF energy.

\subsection{PuO$_{2}$} 
PuO$_{2}$ has the $5f^{4}$ configuration. The crystal field was measured by Kern and co-workers using inelastic neutron scattering (INS) \cite{Kern1999PRB104}.  Our calculated $\Gamma_{1} \rightarrow \Gamma_{4}$ excitation energy of 97 meV agrees reasonably well with the measured value of 123 meV \cite{Kern1999PRB104} and a previous calculation of 99 meV by Colarieti-Tosti {\it et al.\/}\cite{Colarieti-Tosti2002PRB195102}. Note that the splitting was under-estimated in all the calculations, including this work and Refs.~\onlinecite{Colarieti-Tosti2002PRB195102, Magnani2005PRB54405, Gaigalas2009LJP403}. The non-magnetic $5f^{4}$ ground state $\Gamma_{1}$ with $\mu_{S}=\mu=0$,
\begin{eqnarray*}
\mathrm{\includegraphics[height=6mm]{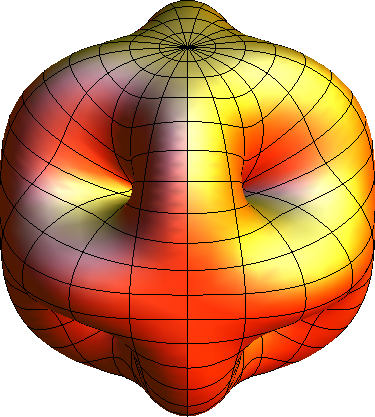}} = 0.7(1,2,3,4)+ 0.32 (1,4,5,6) + 0.32 (2,3,5,6) + \ldots,
\end{eqnarray*}
 is sometimes referred to as four fully filled $\Gamma_{8}$ $f^{1}$ orbitals. We obtained $|\langle \Gamma_{1}| 1,2,3,4 \rangle|^{2}$ = 0.49, showing that such a simplified picture of 4 filled $\Gamma_{8}$ orbitals is not entirely valid and multi-electron correlations account for more than 50\% of the ground state wave function.

\begin{figure}[tbp] 
	\includegraphics[width=0.8 \linewidth]{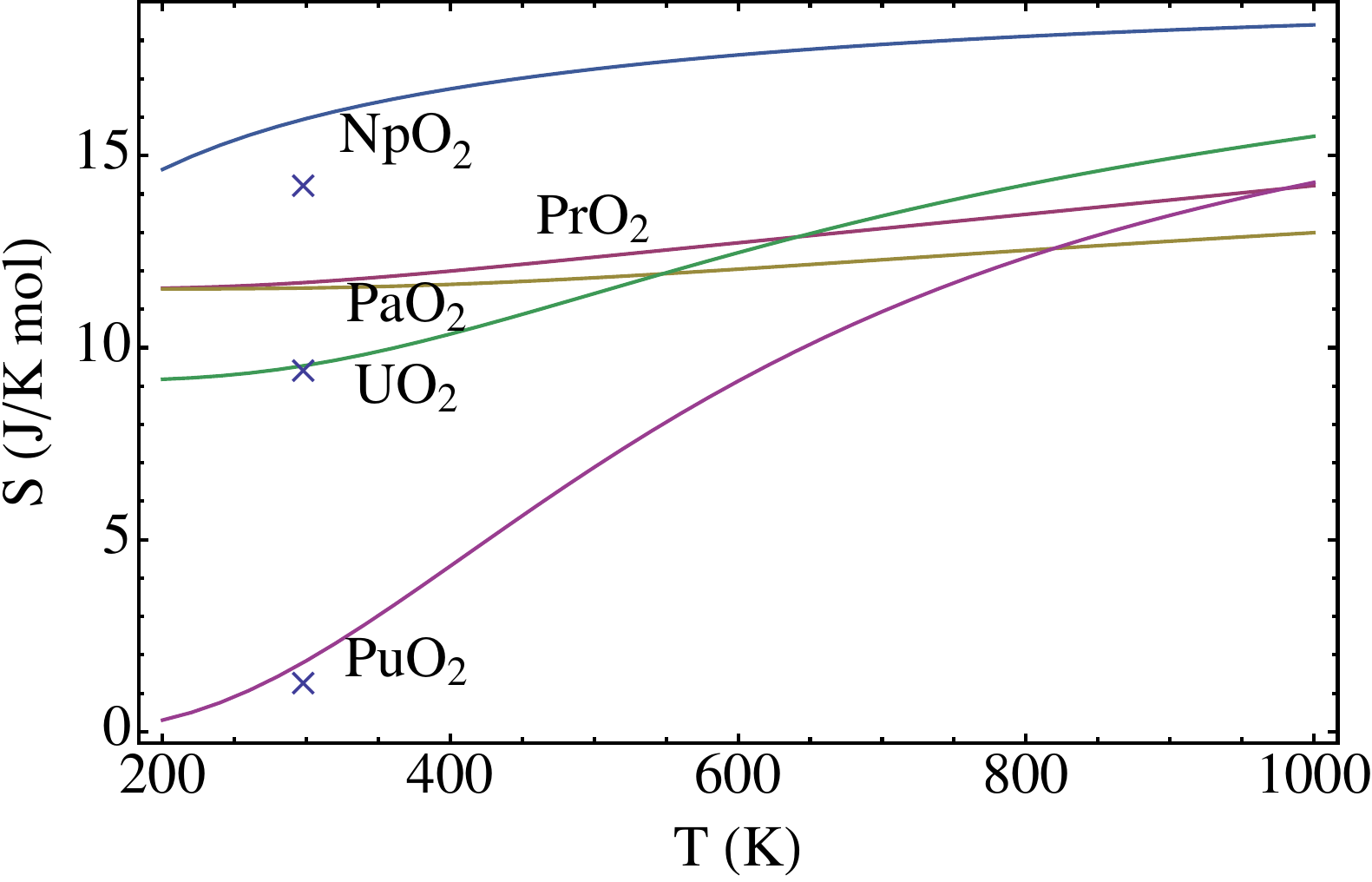} 
	\caption{Predicted electronic entropy of MO$_{2}$ from the calculated CF levels compared to    experimental estimation.} \label{fig:entropy} 
\end{figure}

As a simple application of the CF calculations, Fig.~\ref{fig:entropy} shows the calculated electronic entropy
$$
S_{\mathrm{e}}  = - \sum_{i} p_{i} \ln p_{i},
$$
where $p_{i}= e^{-E_{i}/k_{B}T}/ \sum_{j} e^{-E_{j}/k_{B}T}$ is the Boltzmann probability of the electronic eigenstate $i$. As shown by Konings,\cite{Konings2004JCT121} the vibrational contribution to the total entropy of actinide oxides varies smoothly across the elemental series, while electronic contributions, which depend delicately on the CF excitation energies, cannot be interpolated over the series. To accurately predict thermodynamic properties of actinide oxides, the electronic entropy cannot be ignored. Our predicted $S_{\mathrm{e}}$ (solid curves) agree reasonably well with the results of Ref.~\onlinecite{Konings2004JCT121}  (crosses) at $T=298.15$ K.

In conclusion, we have calculated the CF levels of PrO$_{2}$, PaO$_{2}$, NpO$_{2}$, and PuO$_{2}$. The $f$-electron charge density and on-site correlations are calculated fully self-consistently within a version of LDA+$U$ that removes orbital-dependent self-interaction energies. Good agreement with experimental CF levels and a consistent trend across the actinide series have been achieved. In both NpO$_{2}$ and PuO$_{2}$, substantial contributions of the $\Gamma_{7}$ one-electron excited state are found in the multi-electron crystal field ground states. 

This work was supported by the U.S. Department of Energy, Nuclear Energy Research Initiative Consortium (NERI-C) under grant No.\ DE-FG07-07ID14893, and used resources of the National Energy Research Scientific Computing Center, which is supported by the DOE Office of Science under Contract No.\ DE-AC02-05CH11231.

%

\end{document}